# On Programs and Genomes[1]

Eric Werner[2]

Department Physiology, Anatomy and Genetics
University of Oxford
eric.werner@dpag.ox.ac.uk

**Abstract**

*We outline the global control architecture of genomes. A theory of genomic control information is presented. The concept of a developmental control network called a cene (for control gene) is introduced. We distinguish parts-genes from control genes or cenes. Cenes are interpreted and executed by the cell and, thereby, direct cell actions including communication, growth, division, differentiation and multi-cellular development. The cenome is the global developmental control network in the genome. The cenome is also a cene that consists of interlinked sub-cenes that guide the ontogeny of the organism. The complexity of organisms is linked to the complexity of the cenome. The relevance to ontogeny and evolution is mentioned. We introduce the concept of a universal cell and a universal genome.*

---

[1] This a slightly extended version of Part I of a position paper distributed on November 18, 2007 to the participants of our Balliol Seminar on the Conceptual Foundations of Systems Biology. Based on the first and second part of this paper, I presented my ideas on the global control architecture of genomes. Denis Noble and myself started the seminar in the Michaelmas term in the autumn of 2006 at Balliol College, University of Oxford.

[2] Other current affiliations: Department of Computer Science, University of Oxford; Oxford Advanced Research Foundation (www.oarf.org ); Cellnomica ( www.cellnomica.com ).

1. With special thanks to Denis Noble for motivation and discussions[1].
2. We distinguish **three kinds of information**:
    a. **State information** which gives information about the state of the world.
        i. For example: "Mary is in the house."
    b. **Strategic information** consists of control information which directs the actions of an agent.
        i. Example: "Please, go into the house!"
    c. **Evaluative information** which gives information about utilities or evaluations of a situation.
        i. Example: "Mary likes the house."
    d. These are interrelated. Strategic information depends on its existence on state information. Evaluative information helps construct and choose strategies by way of the formation of goals.

3. **Interpreter-Executive System *IES***. It is good to separate the changing control information in the genome G from the *interpreter and executive system* (*IES*). An interpreter-executive system interprets the control information in the genome and executes that information to generate intra-cellular and external cellular actions.

4. The **separation of control information** from the IES makes possible:
    a. The evolution of organisms.
    b. The variability within and between species.
    c. Without such a separation the egg would have to reflect the complexity of the organism. It would become a kind of regulatory homunculus (see below).
    d. A complex egg would make evolution virtually impossible and hinder the variability of within and between species.

5. Possible directives become active through the IES.

6. **The complexity conservation principle CCP** implies the complexity of development is less than the complexity of the genome G plus the complexity of the IES and environment (Env)[2].
$$K(Dev) <= K(G) \otimes K(IES) \otimes K(Env)$$

7. **Hypothesis**. There exists a mapping from control information in the genome to events in development.

8. **Programs form branching networks**. Programs are not usually just linear systems of commands. They may contain branches and they may contain loops that repeat commands indefinitely or until some condition is met.
    a. A programs action can be conditional on external or internal events.
    b. Hence, programs form implicit branching networks.

9.  **Programs are self controlling**.  They utilize the IES to control what the next step will be.

10. **Programs, viewed as networks** that conditionally control action (through the IES), are also strategies.  And G + IES can be viewed as the controlling core of a **strategic social agent** [3, 4].  In addition to actions that influence the outside world there are communicative acts whose main function is to inform or to direct other social agents.

11. **Cells as social agents.** Cells may thus be viewed as strategic social agents.  Strategic agents SA can be social and form social structures enabling cooperation and competition.

12. **Programs are not just databases**.  A database implies an external agent controls access to information choosing at the will of that external agent.  Thus it presupposes a kind of *regulatory homunculus* that determines how and when the database is to be used.  The database is a passive participant or vesicle from which information is taken or stored.  In fact, the genome contains information that controls its own access.  This may be direct or conditioned on the given external and/or internal situation at a given time.

13. **Multi-agent systems** are distributed systems of strategic agents that coordinate their activity through communication.
    a.  Each agent is controlled by its own strategy or program.
    b.  In other words the programs are distributed over a set of agents.
    c.  Even if the copies of the distributed programs or more generally strategies are identical the activation state of the programs will in general be different.
    d.  Hence, each agent will react based on the particular subprograms (strategies or networks) that are locally active in that agent.

14. **Programs, networks and strategies** are different ways of talking about the organization of control information such as directives or commands that are the input to an interpreter-executive system or IES.

15. **No evolution.** If the complexity of development were in the cell, there would be no evolution and no explanation of the variability between organisms. Otherwise, the egg would have to reflect the complexity of the organism and its development.

16. There is a clear difference between execution of information in the cell and the control information within the genome.

17. There are **different levels of control**. There is control of a cell's external and internal actions. There is control of the cell's interpreter-executive system IES in its coordination with the genome.

18. **Self-controlling genomes.** The crucial point to realize is that the genome is to a large extent self-controlling; the genome itself contains the directive information for its own control, e.g., such as what is activated next based on its control state together with the non-genomic signals that the cell may receive. The network controlling the IES may also be constituted (reconstructed) at each cell division using the IES itself) by meta-information in the genome.

19. **Meta-control.** So, in other words, there is direct control information that directs cell actions and there is information that directs the internal IES network as to which cell control information to activate and then execute next.

20. **Control in Robots**. There are similar control mechanisms in robotics where the computational state changes are driven by the program itself and these state changes in turn activate external control information or directives that are then executed by the controlling executive of the robot. This distinction between activation and externally directed commands allows for more flexibility in the choices available to a robot based on its situation.
In a robot the computer and its program (as hardware or software) are couched inside of it. The robot responds with actions to certain interface-memory states that are controlled by the program. In other words, the program communicates with the robot body via an interface consisting usually of registers (memory units). The robot responds the program's output and the program reacts to the sensory input from the robot. The program does not completely specify or cause the actions by itself. It relies on the robot agent interpreting its commands by actions.
The environment and the interpreting agent-robot interact with the program to give the result. In this sense the program is part of a multi-agent interaction[3]. And yet, this does not mean the program is passive or just an equal player in the formation of the resulting space-time event. It is the control information in the program or a genome that is the essential distinguishing factor in the resulting event (depending upon the complexity of the program, agent, and the environment. It distinguishes a human from a rhino, a flower from a rabbit.

21. **Internal and External Control**. Hence, we distinguish internal control from control of the external actions of an agent. Internal control controls the subsystem responsible for external control. Associated with each type of control is control information that is explicitly or implicitly represented in the agent. The locality of the control information is not the central issue when dealing with the generation of some event. The first constraint is the principle that the event's complexity (at an appropriate level) must reflect the complexity of the generating control information [2]. Given a complex event this then forces a certain complexity on the control information. Then when analyzing an agent responsible for the generation of the event, we can look for what structures could possibly contain this complex control information. In the case that the agent is a zygote or fertilized egg, the information to generate a complex organism most naturally lies in its diploid genome. The egg itself contains the interpreter-executive system IES but not the actual control information to generate the organism.

22. **Communication** between agents such as animals, robots or cells can be used to constrain strategic possibility spaces (and thereby strategies) so enable coordination of activity. Indeed, there may even be linguistic or signaling strategies, such as protocols, that underlie and coordinate such communications [3, 4].

23. **Information Poor Signals with Rich Interpretations.** When a cell receives a signal and that signal ultimately activates a whole genomic network, it is the activated network that is rich in information content, not the signal itself. Indeed, the complex potential response to a signal by the interpreting agent is what gives the signal its pragmatic meaning. Thus, a signal itself will in general have little relevant information content. Rather the interpretation of the signal by the receiver contains the complexity, its rich information content. Note, in the cell even the IES need not be very complex in order to give a complex interpretation to a signal. Rather, all the IES needs to do is map the signal to the appropriate area of the genome and then use the information in that local genome network to generate a full, rich and complex interpretation.

24. **Algorithms** that generate "rational" strategic responses to situations, based on some evaluative criteria, are implicit reactive strategies or generate a strategy on the fly.

25. **What are genes?** Clearly if we think of Mendel's original idea of genes as units that determine global attributes of living organisms, then genes cannot just be proteins which are mostly just parts that make up a structure. Just as we distinguish the architecture of a building from its parts, as a church and a skyscraper can be very different yet be constructed from the same basic parts, so we distinguish the genes as parts from those sequences that underlie the generative control architecture of the genome. Rather it may better to think of genes in Mendel's sense as genomic regulatory sub-networks that control aspects of development. Viewed this way genes can be combined to form larger networks.

26. **Cenes**. For the last several years in talks, I have called such sub-networks *cenes* or control genes networks to indicate their regulatory and constructive genetic function. The smallest cene is just a network of one node, a control gene, which controls an individual action. Because cenes are networks and not just protein parts, they have a flexibility that allows for evolution of the control architecture of the genome. The part of the genome underlies the control of development of the whole organism is just a linked collection of cenes. In other words, this part of the genome is just a large cene or cenome.

27. **Cenomes**. The *cenome* is the global regulatory network that controls development. This is not to say that physics is not important in development, it is. However, the generative control information that determines the structural and dynamic complexity of the four dimensional event we call embryonic development is mostly contained in the cenome. The compositionality permitted by cenes provides a modularity of control that we do not have with the molecular view of genes as coding for proteins.

28. **Cell Communication Strategies**. Viewed as proto-strategies, cenes form the basis of cell social life by specifying directly or indirectly the strategies that are necessary for the cell to be a social agent in a vast multiagent system. *Communication cenes* form the basis of protocols that coordinate inter-cellular cooperative social action.

29. **Information can exist independently at different levels of ontology**. Consider a sequence of possibly different books that all have either a red or white cover. We can construct a sequence of n such red and white books to represent any on the 2^n possible binary sequences. Hence, we can represent any binary number. The content of the books we assume is independent of its color. Let us assume, the content may consist of any English text. Clearly the information given by the sequence of books is totally independent of their contents. However, we could construct a sequence of some length n, using book with k different contents, such that that sequence of red and white books gives the same information as the content of the k books types used to construct the sequence. Indeed, we may have only one book content, but with two covers, red and white, such that the content of the book is used by an interpretive-executive agent to construct the sequence of books that contains the information of their content.
    The point I am making is that there can be levels on ontology such that the information at one level can be totally independent of another level. However, there can also be interactions between levels of ontology and information.

30. **Downward Causation**. We have seen how information can flow upwards from book content to book sequence. We can also have a kind of "downward causation" via a downward flow of information when a sequence of n red and white books is used by an interpreter-executive agent to construct the contents of a new red or white book.
    The complexity conservation principle regulates what information can in principle be created when information flows within or between levels of ontology. Interactionism, where relatively simple agents and their local interactions determine global structure or information output severely limits the complexity of that output. Hence, my skepticism of Turing's method[5], and of Wolpert's positional information approach[6]. The complexity of the 4D-organism's development just does not seem to be captured by their relatively simple informational bases. Indeed, they entirely ignore the information in the genome.

31. **Mapping Relations between Levels of Ontology and Information**. A level $L_n$ is causally *independent* of level $L_m$ if there is no mapping (deterministic or probabilistic, partial or complete) from information at $L_m$ to information at $L_n$. For example, the properties and information in the bases that make up DNA give no information about the information in the sequence that makes up DNA. The bases are *information poor* compared to the information in a DNA sequence. However, when we look at the cell, the information in the Genome sequence is *information rich* when compared to the information in the cell. There may be mappings from an information poor level to an information rich level if the IES that does the mapping contains the relevant information of the rich information it is generating. Another way this can occur if the information rich area is indexed by addresses. Then the mapping relation created by the IES can relate the information poor source with an information rich receiver by using the indices to do the mapping. Then the rich information need not reside in the IES.

32. **The Role of Physics**. The hypothesis that most of the control information responsible for the complexity of organisms lies in the genome, does not preclude a central and formative role for cell interactions based on their physics and chemistry. Nor does it contradict the notion that development is a step by step process that utilizes the given state of the system to determine what happens next [7]. However, it is a matter of weight- what information is more relevant to the unique organism in question. Clearly the physics of cell-cell interaction will be quite similar for mammals if not for plants. If so then the physics alone cannot be controlling factor responsible for the differences seen in mammals.

33. **A Universal Egg**. If it turns out that there is the possibility of constructing a universal egg that forms a zygote when inserting a diploid genome plus its initial boot system information, then clearly the genome is the central player in development and not the egg.

34. **A thought experiment: The Universal Cell UC** Imagine a universal cell UC and a universal maternal matrix UM that is such that when we add to UC a initial genome $ig_O$ then UC plus $ig_O$ in UM generates the initial egg state E of the organism O. Then, if we add the genome $G_O$ to E (where UC $\otimes$ $ig_O$ $\rightarrow$ E) then the combination $G_O$ $\otimes$ E will generate the organism O given appropriate maternal matrix MM such as a womb or external egg. Hence, the only unique information contributed by the resulting egg is $ig_O$. Thus, if we start with a universal cell we can mimic any organism O given $ig_O$;$G_O$. Granted, the maternal matrix is important as a necessary condition and the result of $ig_O$;$G_O$ may be different in a different maternal matrix. Still the informational contribution from MM will not be the information that generates O. This thought experiment, while presently unrealistic, shows some fundamental principles about organisms and their genomes. The development of the organism can be viewed as multi-agent interaction between UC, the program $ig_O$;$G_O$, the maternal matrix MM and the environment $\Omega$. Using the symbol $\otimes$ for interaction, $\rightarrow$ for generates, we can state this concisely as a formula: UC $\otimes$ $ig_O$;$G_O$ $\otimes$ MM $\otimes$ $\Omega$ $\rightarrow$ O. The relevant information/complexity to generate the organism O is contained in the genome $G_O$, and not in the unique information $ig_O$ in the organism's egg cell E, not in the maternal matrix MM, and not in the environment $\Omega$.

The universal cell UC with initiating information $ig_O$ generates the egg E, UC[$ig_O$] $\rightarrow$ E. The egg with $G_O$ generates O, E[$G_O$] $\rightarrow$ O (given MM and $\Omega$). Or, another way of putting it: UC[$ig_O$;$G_O$] $\rightarrow$ O. Putting in the maternal matrix and the environment $\Omega$ explicitly:
UC[$ig_O$;$G_O$] $\otimes$ MM $\otimes$ $\Omega$ $\rightarrow$ O.

35. **Universal cells and Universal Turing Machines.** Universal cells are in many ways analogous to Universal Turing machines. A Universal Turing machine can mimic any Turing machine given an appropriate initiating program. Turing invented a mathematical analogue of a computer, now called a Turing Machine, which was capable of executing programs in form of binary code [8]. Simplifying a bit, a Turing Machine TM consists of a tape with data, a set internal states, and a transition function that transforms a state given some input data on its tape, into a new state and possibly puts some new output data on the tape. This output data can then be used again as input. TM $\otimes$ state $\otimes$ data $\rightarrow$ new state and output. A universal Turing machine UTM can interpret any program pTM that when put on the tape prior to the data, will result in the UTM mimicking any Turing Machine TM described by pTM. So, UTM $\otimes$ pTM;data $\rightarrow$ output. For any data, UTM $\otimes$ pTM; data is output equivalent to TM $\otimes$ data. What I am suggesting is that we apply these concepts to cells. Then a universal cell UC can mimic any cell if we first add an initiating genome $ig_O$ that converts UC into an egg cell that can generate organism O given genome $G_O$.

36. **Minimal cell versus a universal cells**. There are efforts to create a minimal cell. A minimal cell is a cell has been defined as a cell with smallest number genes that can survive on its own. More generally, since genes by themselves must be regulated, a minimal cell is a cell that has the shortest length genome that can survive. If we take the length of the minimal genome as a measure of complexity of the cell then a minimal cell has the least complexity. It is clear that there it is logically possible that there can be more than one minimal cell in both senses of minimal.

    A universal egg cell has the additional capability to generate any organism O given the initiation genome $ig_O$ plus the genome $G_O$ of the organism (given a maternal matrix $MM_O$ for that organism O). Stem cells have some of the properties of universal cells. They have the capacity to differentiate into any cell type given the appropriate initial conditions. In cloning (somatic cell nuclear transfer) we take the nucleus of a cell, insert it into an enucleated egg to give a zygote equivalent to a fertilized ovum that can then generate an organism, given the appropriate maternal matrix.

    A universal egg cell has a nucleus but not the genome and not the specific transcription factors of zygote. A universal egg cell transforms into a particular zygotic equivalent through the execution of the initiation genome $ig_O$.

37. **Maternal TF's.** The egg contains maternal proteins and transcription factors not produced by the egg's genome -though they are produced by the mother using her genome. If a genome $G_A$ of organism A in an egg (in the context of the a maternal matrix such as a womb) results in the embryogenesis of the organism A and another genome $G_B$ of an organism B in an identical egg (given an identical maternal matrix) generates B, then the unique features distinguishing A and B cannot be due to the information in the egg nor in the maternal matrix. Rather, the relevant information needed to generate the organism lies in the genome. For example, the fact that two brothers can be so different appears to be due to the difference in their genomes and not due to differences in the eggs containing their genomes.

38. **Dinosaur versus chicken.** ( See [9]) Or, if we would insert a dinosaur's DNA into a chicken egg, and it developed into a dinosaur, we would have to conclude that the DNA and not the information in the egg was responsible its development as a dinosaur and not a chicken. True we would need maternal dinosaur transcription factors to bootstrap the process of development of the dinosaur. And, there may be other factors such as the size, nutrients, and thickness of the egg shell that are relevant. But it is clear that they do not contain the relevant control information needed to generate the dinosaur. While this is a thought experiment verging on science fiction, it makes the point that the genome and not the egg or the maternal matrix contains the relevant information needed to generate an organism. For Noble, this process, not being controlled by any 'program' is the result of a mysterious interaction of equal partners. I am just saying they are far from equal in information content.

39. **Universal Genome.** There is a caveat; one could imagine a *universal genome UG* that contained all genomes of all organisms. Then the initial state of the activating transcription factors may be all that distinguishes one organism from another. However, for most organisms the more reasonable hypothesis is that the genome contains the information for an organism's morphological and functional development and not the egg nor the maternal context. Furthermore, even if there were just one universal genome then the differences in organisms would be due to differences in sub-genomes of **UG** where the role of transcription factors is restricted to activating or deactivating these sub-genomes. Thus, the distinguishing information is still unique to an organisms activated sub-genome.

40. **Universal Genome in a Universal Cell.** Given a universal genome UG in a universal cell UC then the resulting organism would be totally determined by the initiation information $ig_o$. Recall $UC[ig_o] \rightarrow E$. And $E[G_o] \rightarrow O$ (given MM and $\Omega$). Hence, $UC[ig_o; G_o] \rightarrow O$. Thus, $UC[ig_o; G_o] \otimes MM \otimes \Omega \rightarrow O$. $UC[ig_o; ip; UG] \otimes MM \otimes \Omega \rightarrow O$. We need the additional initiation process ip to translate the $ig_o$ egg state to one compatible the universal genome. For there may be addressing conflicts when we combine all possible genomes. When constructing a universal genome from genomes $G_1, \ldots, G_n$ we have to have to add a new promoter region for each genome to be able to activate that particular genome and not the others. Either we have to deactivate, say by methylation, the other non-used genomes or we have to change all the promoter regions so that they do not conflict with each other and do not have unwanted interactions. The initial activating transcription factors then activate the unique genome $G_i$ and not any of the others. In fact, $ig_o$ for organism $O = O_i$ might activate several genomes in UG if UG is not first made coherent and $ig_o$ is not translated appropriately to activate only the particular genome responsible for generating $O_i$.

41. **Limits of Interactionism.** Given a set of n redundant agents $P_i$ all the same then these agents cannot generate an event (structure) more complex than the complexity of an agent $P_i$ plus the complexity of the number n plus some constant c. **Argument**. If they could generate a structure S such that K(S) has a complexity greater than K($P_i$) + K(n) + c, then we could construct a program X to generate $P_i$ n times and their interactions would generate S. However, then we have a program that generates S that has a length shorter than K(S), contradicting our assumption that K(S) is more complex than the agents $P_i$.

42. **Random Information.** Given some stochastic source that generates complex, but random information, this information will, in general, not be useful in generating a particular structure. Indeed, such sources of randomness have to be avoided for the process of development to work. The couching of the embryo in a safe maternal environment put a boundary between the embryo and a relatively random environment that generates irrelevant information for the embryo's development. Looking internally at the detailed molecular state of the egg is to a large extent irrelevant and would be harmful to the developmental process if it is not filtered out or ignored by the IES. Thus, most of the non-genomic information in the egg is functionally irrelevant.

43. **Temporal Developmental Control Information.** As the organism develops the state of the IES of each cell together with its interactive state with the genome G changes. In general the states of the cells in the organism will differ. While the genome may remain the same, the IES state will often differ. For each $t \in \Psi_O$ there exists information in $O_t \otimes G_t$ such that $O_t \otimes G_t \otimes E \in O_{t+1}$
Critique: Each cell in $O_t$ may be in a different state. If we include the genome's epigenetic state (e.g., methylation), in its overall state at some time t, then the genome's state $G_{c,t}$ is relativized to the cell c at time t. The state of the $IES_c$ is relativized to the cell c. In general, for any two cells c1, c2 in $O_t$, c1[$IES_{c1,t}$, $G_{c1,t}$] ≠ c2[$IES_{c2,t}$, $G_{c2,t}$]

44. **Gene Expression State and the Cenome.** The gene expression state of a cell must somehow be controlled as to when and where it occurs. Hence, for each possible gene expression state there must exist control information in the genome to activate, transform and deactivate that state. Even if the state is the result of cell to cell signaling, the cell's rich response to any signal demands that it contain sufficient control information to enable the activation of that rich response. Furthermore, it must contain the information that generates the rich response itself. Thus, we have both control of activation and the response to a signal. Both must somehow be represented in the cell. In general, due to the complexity and changing dynamics this control information is most likely in the genome. Hence, the cenome will contain much of the complexity required for the control of gene expression states.